\documentclass{article}

\usepackage{amsmath, epsfig, citesort}
\usepackage{amssymb}
\usepackage{amsfonts}
\usepackage{latexsym}
\usepackage{longtable}
\usepackage{tabularx}

\newtheorem{theorem}{Theorem}
\newtheorem{prop}[theorem]{Proposition}

\newtheorem{corollary}[theorem]{Corollary}

\newtheorem{lemma}[theorem]{Lemma}

\newcommand{\qed}{{\hfill\rule{4pt}{7pt}}}
\def\pf{\noindent {\it Proof.} }

\begin{document}
\title{New bounds of permutation codes \\under Hamming metric and Kendall's $\tau$-metric}

\author{Xin Wang$^{\text{a,b}}$, Yiwei Zhang$^{\text{a}}$, Yiting Yang$^{\text{c}}$ and Gennian Ge$^{\text{a,d,}}$\thanks{Corresponding author (e-mail: gnge@zju.edu.cn). Research supported by the National Natural Science Foundation of China under Grant Nos. 11431003 and 61571310.}\\
\footnotesize $^{\text{a}}$ School of Mathematical Sciences, Capital Normal University, Beijing 100048, China\\
\footnotesize $^{\text{b}}$ School of Mathematical Sciences, Zhejiang University, Hangzhou 310027, Zhejiang, China\\
\footnotesize $^{\text{c}}$ Department of Mathematics, Tongji University, Shanghai 200092, China\\
\footnotesize $^{\text{d}}$ Beijing Center for Mathematics and Information Interdisciplinary Sciences, Beijing 100048, China.\\}

\date{}\maketitle

\begin{abstract}
Permutation codes are widely studied objects due to their numerous applications in various areas, such as power line communications, block ciphers, and the rank modulation scheme for flash memories. Several kinds of metrics are considered for permutation codes according to their specific applications. This paper concerns some improvements on the bounds of permutation codes under Hamming metric and Kendall's $\tau$-metric respectively, using mainly a graph coloring approach. Specifically, under Hamming metric, we improve the Gilbert-Varshamov bound asymptotically by a factor $n$, when the minimum Hamming distance $d$ is fixed and the code length $n$ goes to infinity. Under Kendall's $\tau$-metric, we narrow the gap between the known lower bounds and upper bounds. Besides, we also obtain some sporadic results under Kendall's $\tau$-metric for small parameters.
\medskip


\smallskip


\smallskip
\end{abstract}

\section{Introduction}

Let $S_n$ be the symmetric group on $n$ elements. A permutation code is a subset of $S_n$ satisfying certain constraints. Permutation codes have been studied under various metrics according to specific applications. 
In this paper we focus on two kinds of metrics, the Hamming metric and the Kendall's $\tau$-metric. We now briefly introduce the motivations for these two metrics.

During the last decade, permutation codes under Hamming metric have attracted considerable attention, due to their applications in data transmission over power lines \cite{FV,PVYH,V}.
In the power line application, there are three main forms of noise which affect the transmission: the permanent narrow-band noise, the impulse noise of short duration and white Gaussian noise (background noise). In many traditional data transmission media (e.g., telephone lines and satellite communications), white Gaussian noise is the dominant type of error affecting the system. However, the other two types of errors play important roles in the power line application. In \cite{FV} and \cite{V}, permutation codes under Hamming metric are used to correct errors for this type of transmission. Besides, permutation codes under Hamming metric have been applied in the design of block ciphers \cite{DCL}.

The research on permutation codes under Kendall's $\tau$-metric
has a relatively shorter history, which
originates from the development of flash memories. Flash memory incorporates a set of cells maintained at a set of levels of charge to encode information. The chief disadvantage of flash memories is their inherent asymmetry between cell programming (injecting cells with charge) and cell erasing (removing charge from cells). While raising the charge level of a cell is an easy operation, reducing the charge level from a single cell would require completely erasing a whole large block to which the cell belongs and then reprogramming, which will limit the lifetime of a flash memory. Therefore, over-programming (increasing charge level on a cell above the desired amount) is a severe problem. Moreover, flash memories meet common errors due to charge leakage and reading disturbance. In order to overcome these problems, the novel framework of {\it rank modulation} is introduced in \cite{Jiang}. Instead of encoding information with the absolute values of charge levels, data is represented by the relative rankings of the charge levels on a group of cells. That is, if we have $n$ cells and $c_1,c_2,\dots,c_n \in \mathbb{R}$ represent the charge levels, then this group of cells is said to encode the permutation $\sigma\in S_n$ such that $c_{\sigma(1)} > c_{\sigma(2)} > \dots > c_{\sigma(n)}$. In this framework, we save us the trouble to deal with errors which only slightly affect the absolute values of charge levels but do not affect the relative rankings. However, sometimes the errors in the charge levels may be large enough to cause some disturbance in the relative rankings. To detect and/or correct such errors we need an appropriate distance measure. Several metrics on permutations are used for this purpose such as Kendall's $\tau$-metric \cite{Jiang2,Barg,Mazumdar,Buzaglo2}, Ulam metric \cite{Farnoud} and $l_\infty$-metric \cite{Klove,Tamo}.

The rest of this paper is organized as follows. In Section \ref{preliminary} we give the definitions and notations of permutation codes under Hamming metric and Kendall's $\tau$-metric and summarize some important known facts regarding the bounds, and then we introduce the corresponding graph models as a preparatory step for the upcoming analysis. A lower bound of permutation codes under Hamming metric is given in Section \ref{hamming} which improves the Gilbert-Varshamov bound by a factor of $n$. A lower bound of permutation codes under Kendall's $\tau$-metric is given in Section \ref{kendall}. In Section \ref{sporadic} some other sporadic results concerning permutation codes under Kendall's $\tau$-metric are listed. We conclude in Section \ref{conclusion}.

\section{Preliminaries} \label{preliminary}

In this section we first give some definitions and notations for permutation codes under Hamming metric and Kendall's $\tau$-metric and summarize some important known facts regarding the bounds. 

Let $[n]$ denote $\{1,2,\dots,n\}$. Let $\pi=[\pi_1,\pi_2,\dots,\pi_n]$ be a permutation over $[n]$ such that for each $i\in[n]$ we have $\pi(i)=\pi_i$. This form is known as the {\it vector notation} for a permutation. For an integer $x\in[n]$, $\pi^{-1}(x)$ indicates the position of $x$ appearing in $\pi$. For two permutations $\sigma$ and $\pi$, their composition, denoted by $\sigma\pi$, is the permutation with $\sigma\pi(i)=\sigma(\pi(i))$ for all $i\in[n]$. All the permutations under this operation form the noncommutative group $S_n$ known as the symmetric group on $[n]$ of size $|S_n|=n!$. Denote by $\varepsilon\triangleq[1,2,\dots,n]$ the identity element of the group. For an unordered pair of distinct numbers $x,y\in[n]$, this pair forms an {\it inversion} in a permutation $\pi$ if $x<y$ and simultaneously $\pi^{-1}(x)>\pi^{-1}(y)$. Let $I(\pi)$ denote the total number of inversions in a permutation $\pi$. $\pi$ is called an even/odd permutation accordingly due to the parity of $I(\pi)$.


\subsection{Hamming metric}

For two permutations $\sigma$ and $\pi$, the {\it Hamming distance} between them is the number of positions in which their vector notations differ, i.e.
\begin{equation*}
d_H(\sigma,\pi)=|\{i \in [n]:\sigma_i\neq \pi_i)\}|.
\end{equation*}

For $1\le d \le n$, we say that $\mathcal{C}\subset S_n$ is an {\it$(n,d)$-permutation code under Hamming metric}, if $d_{H}(\sigma,\pi)\ge d$ for every two distinct permutations $\sigma,\pi\in \mathcal{C}$. Denote the largest size of an $(n,d)$-permutation code under Hamming metric as $A_H(n,d)$ and a code attaining this size is said to be optimal. The exact value of $A_H(n,d)$ and the constructions of optimal codes are the main research objectives.
There are some fundamental results by basic combinatorial techniques.

\begin{prop}{\rm\cite[Proposition 1.1]{CCD}}
  \begin{enumerate}
    \item $A_H(n,2) = n!$;
    \item $A_H(n,3) = n! / 2$;
    \item $A_H(n,n) = n$;
    \item $A_H(n,d) \le n A_H(n-1,d)$;
    \item $A_H(n,d) \le n! / (d-1)!$.
  \end{enumerate}
\end{prop}

However, deciding $A_H(n,d)$ turns out to be difficult for $4\leq d\leq n-1$, except for some specifical cases.
\begin{prop}
\begin{enumerate}
  \item {\rm\cite{CKL}} If there are $m$ mutually orthogonal Latin squares of order $n$, then $A_{H}(n,n-1)\geq mn$. In particular, if $q$ is a prime power, then $A_H(q,q-1)=q(q-1)$.
  \item {\rm\cite{FD}} If $q$ is a prime power, then $A_H(q+1,q-1)=(q+1)q(q-1)$.
\end{enumerate}
\end{prop}

We now summarize some important general results concerning the lower and upper bounds of $A_H(n,d)$.

Let $D(n,k)$ ($k=0,1,\dots,n$) denote the set of all permutations in $S_n$ which are exactly at distance $k$ under Hamming metric from the identity permutation $\varepsilon$:
$$D(n, k) = \{ \pi \in S_n: d_H (\pi, \varepsilon) = k \}.$$

A {\it derangement} of order $k$ is a permutation $\pi\in S_k$ with no fixed points, i.e., $\pi_i\neq i$ for $1\le i \le k$. Let $D_k$ be the number of derangements of order $k$. Then the cardinality of $D(n,k)$ is
\[|D(n,k)| = {n \choose k} D_k.\]

For any permutation $\pi\in S_n$, the {\it Hamming ball} of radius $r$ centered at $\pi$, denoted as $B_H(\pi,r)$, is defined by $B_H(\pi,r)\triangleq \{\sigma\in S_n: d_H(\sigma,\pi)\le r\}$. Clearly under Hamming metric the size of a ball of radius $r$ does not depend on the center of the ball and we denote its size as $B_H(r)$:
\[
B_H(r)=\sum_{k=0}^{r}|D(n,k)|.
\]

The Gilbert-Varshamov bound and sphere-packing bound for permutation codes under Hamming metric are well known.

\begin{prop}
\[
\frac{n!}{B_H(d-1)}\leq A_H(n,d)\leq \frac{n!}{B_H(\lfloor\frac{d-1}{2}\rfloor)}.
\]
\end{prop}

Improved lower bound for the case when $d$ is fixed and $n\rightarrow \infty$ is derived by Gao, Yang and Ge in \cite{GYG}.

\begin{prop}{\rm\cite[Theorem 10]{GYG}} \label{GYG}
Let $d$ be fixed and $n\rightarrow \infty$. Then
\[
A_H(n,d)\geq \Omega(\log n \frac{n!}{B_H(d-1)}).
\]
\end{prop}

Later, Tait, Vardy and Verstra\"{e}te \cite{TVV} consider the case when the ratio $d/n$ is fixed and improve the Gilbert-Varshamov bound by a factor of $n$.
\begin{prop}{\rm\cite[Theorem 2]{TVV}} \label{TVV}
Let $d/n$ be a fixed ratio with $0<d/n<0.5$. Then as $n\rightarrow\infty$, we have
\[
A_H(n,d)\geq \Omega(n \frac{n!}{B_H(d-1)}).
\]
\end{prop}


\subsection{Kendall's $\tau$-metric}

Given a permutation $\pi=[\pi_1,\pi_2,\dots,\pi_n]\in S_n$, an {\it adjacent transposition} is an exchange of two adjacent elements $\pi_i,\pi_{i+1}$, for some $1\le i \le n-1$, resulting in the permutation $[\pi_1,\dots,\pi_{i-1},\pi_{i+1},\pi_i,\pi_{i+2},\dots,\pi_n]$. The {\it Kendall's $\tau$-distance} between two permutations $\sigma$ and $\pi$, denoted by $d_K(\sigma,\pi)$, is the minimum number of adjacent transpositions required to transform one permutation into the other. For example, the Kendall's $\tau$-distance between $\pi_1=[1,2,3,4,5]$ and $\pi_2=[2,3,1,5,4]$ is three, since we may do the adjacent transpositions $[1,2,3,4,5]\rightarrow[2,1,3,4,5]\rightarrow[2,3,1,4,5]\rightarrow[2,3,1,5,4]$ and one may easily check that only two adjacent transpositions are not enough. A well-known equivalent expression for $d_K(\sigma,\pi)$ \cite{Jiang2} is as follows:

\begin{equation*}
d_K(\sigma,\pi)=|\{ (i,j): \sigma^{-1}(i)<\sigma^{-1}(j) \wedge \pi^{-1}(i)>\pi^{-1}(j)\}|.
\end{equation*}

For $1\le d \le {n \choose 2}$, we say that $\mathcal{C}\subset S_n$ is an {\it$(n,d)$-permutation code under Kendall's $\tau$-metric}, if $d_K(\sigma,\pi)\ge d$ for every two distinct permutations $\sigma,\pi\in \mathcal{C}$. Denote the largest size of an $(n,d)$-permutation code under Kendall's $\tau$-metric as $A_K(n,d)$ and a code attaining this size is said to be optimal. The exact value of $A_K(n,d)$ and the constructions of optimal codes are the main research objectives.
There are some fundamental results as follows.

\begin{prop}
  \begin{enumerate}
    \item $A_K(n,2) = n! / 2$ and the optimal codes are either the set of all even permutations or the set of all odd permutations;
    \item {\rm\cite[Theorem 10]{Buzaglo2}} For $\frac{2}{3}{n\choose 2}<d\le{n\choose 2}$, $A_K(n,d) = 2$;
  \end{enumerate}
\end{prop}

However, deciding $A_K(n,d)$ turns out to be difficult for $3\leq d\leq \frac{2}{3}{n\choose 2}$. We now summarize some important results concerning the lower and upper bounds of $A_K(n,d)$.

Similarly as above we first introduce the Gilbert-Varshamov type lower bound and the sphere-packing upper bound.
For any permutation $\pi\in S_n$, the {\it Kendall's $\tau$-ball} of radius $r$ centered at $\pi$, denoted as $B_K(\pi,r)$, is defined by $B_K(\pi,r)\triangleq \{\sigma\in S_n: d_K(\sigma,\pi)\le r\}$. Clearly under Kendall's $\tau$-metric the size of a ball of radius $r$ does not depend on the center of the ball and we denote its size as $B_K(r)$.
The Gilbert-Varshamov bound and sphere-packing bound for permutation codes under Kendall's $\tau$-metric are as follows:
\begin{prop}{\rm\cite[Theorems 17 \& 18]{Jiang2}}
\begin{equation*}
\frac{n!}{B_{K}(d-1)}\le A_k(n,d) \le \frac{n!}{B_K(\lfloor\frac{d-1}{2}\rfloor)}.
\end{equation*}
\end{prop}

For two permutations $\sigma$ and $\pi$ with $d_K(\sigma,\pi)=1$, the {\it double ball of radius $r$} centered at $\sigma$ and $\pi$ is defined by $DB(\sigma,\pi,r)\triangleq B(\sigma,r) \cup B(\pi,r)$. Denote by $DB_{n,r}$ the double ball of radius $r$ in $S_n$ centered at the identity permutation $\varepsilon$ and the permutation $[2,1,3,4,\dots,n]$.
Improved upper bound for the cases when $d$ is even is derived in \cite{Buzaglo2}, using a code-anticode approach.

\begin{prop}{\rm\cite[Corollaries 3 \& 5]{Buzaglo2}} \label{BEEven}
If a code $\mathcal{C}\subset S_n$ has minimum Kendall's $\tau$-distance $d$, and an anticode $\mathcal{A}\subset S_n$ has maximum Kendall's $\tau$-distance $d-1$, then $|\mathcal{C}|\cdot|\mathcal{A}|\le n!$. Particularly, since $DB_{n,r}$ is an anticode of diameter $2r+1$, so we have
\begin{equation*}
A_K(n,2(r+1))\le \frac{n!}{|DB_{n,r}|}.
\end{equation*}
\end{prop}


Regarding the improvements on the lower bound, first we note that we could just concentrate on $A_K(n,d)$ with odd $d$, since we have the following simple but useful fact \cite{Jiang2}:
\begin{lemma}{\rm\cite[Theorem 26]{Jiang2}} \label{even}
For all $n$ and $t\ge1$ we have $A_K(n,2t)\ge\frac{1}{2}A_K(n,2t-1)$.
\end{lemma}

An important improvement of the lower bound is derived in \cite{Barg}, which is a generalization of a construction of an $(n,3)$-permutation code under Kendall's $\tau$-metric using codes in the Lee metric appeared in \cite{Jiang2}. The generalization leads to a construction of an $(n,2t+1)$-permutation code under Kendall's $\tau$-metric, which is of optimal size up to a constant factor, for a fixed $t$.

\begin{prop}{\rm\cite[Theorem 4.5]{Barg}} \label{Barg}
Let $m=((n-2)^{t+1}-3)/(n-3)$, where $n-2$ is a prime power. Then we have
\begin{equation*}
A_K(n,2t+1)\ge
\begin{cases}
n!/(t(t+1)m),& t\textup{ odd};\\
n!/(t(t+2)m),& t\textup{ even}.
\end{cases}
\end{equation*}
\end{prop}


\subsection{Graph models}

Finally in this section we introduce a natural connection between codes and independent sets of their corresponding graphs. A graph $G$ consists of a set of vertices $V(G)$ and a set of edges $E(G)$. Two vertices $u$ and $v$ are called adjacent if $\{u,v\}\in E(G)$. An independent set in a graph is a set of vertices where every pair of vertices are nonadjacent. The size of the largest independent set in a graph $G$ is called the independence number, denoted as $\alpha(G)$.

Let $G_H$ and $G_K$ be graphs with the same vertex set $S_n$. Two vertices are connected in $G_H$ (respectively, $G_K$) if and only if their Hamming distance (respectively, Kendall's $\tau$-distance) is at most $d-1$. Then, an $(n,d)$-permutation code under Hamming metric (respectively, Kendall's $\tau$-metric) is equivalent to an independent set in $G_H$ (respectively, $G_K$). Via this natural connection, graph-theoretic tools for analyzing independence numbers can be used for analyzing bounds of codes.

In this paper, we mainly use a coloring approach to analyze the lower bound of the independence numbers of $G_H$ and $G_K$. A {\it coloring} of a graph assigns a color to each vertex. It is called a {\it proper coloring} if it never assigns the same color to both endpoints of an edge. The chromatic number of a graph $G$, denoted by $\chi(G)$, is the smallest integer $k$ such that a proper coloring of $G$ using $k$ colors exists. Given a proper coloring, by definition every set of vertices with a same color constitutes an independent set. So we have
\begin{lemma} \label{color}
$\alpha(G)\geq|V(G)|/\chi(G)$.
\end{lemma}
Thus, lower bounds of $\alpha(G)$ could be derived via analyzing upper bounds of $\chi(G)$.

Another fact concerning the independence number of a graph is as follows. An {\it automorphism} of a graph $G$ is a bijective function $f:V(G)\rightarrow V(G)$, such that for any pair of vertices $u,v\in V(G)$, $(f(u),f(v))\in E(G)$ if and only if $(u,v)\in E(G)$. A graph $G$ is called {\it vertex transitive} if for any two vertices $u$ and $v$, there exists some automorphism $f:V(G)\rightarrow V(G)$ such that $f(u)=v$. Then it is well known that (see, for example \cite{Godsil})

\begin{lemma} \label{local}
If the graph $G$ is vertex-transitive and $G'$ is any induced subgraph of $G$. Then we have
$$\frac{\alpha(G)}{|V(G)|}\le \frac{\alpha(G')}{|V(G')|}.$$
\end{lemma}

\section{A lower bound of permutation codes under Hamming metric} \label{hamming}

In this section we consider the lower bound of $A_H(n,d)$ by giving a proper coloring for the graph $G_H$.

\begin{theorem}\label{pc}
Let $n,d$ be integers, $4\le d \le n-1$. Let $p$ be the smallest prime number greater than or equal to $n$. Then, we have
\[
A_H(n,d)\geq \frac{n!}{p^{d-2}}.
\]
\end{theorem}

\pf
Let $\mathbb{Z}_p=\mathbb{Z}/p\mathbb{Z}$ denote the residue class modulo $p$. View the vector notation of a permutation as an $n\times1$ vector. Consider the coloring map
\[
f:S_n\rightarrow \mathbb{Z}_p^{d-1},
\]
whose value at $\sigma\in S_n$ is determined by
\[
f(\sigma)=A\sigma~~\pmod{p},
\]
where $A$ is a $(d-1)\times n$ Vandermonde matrix as follows ($x_1,x_2\dots,x_n$ are distinct numbers in $\{0,1,\dots,p-1\}$):
\[\left(
  \begin{array}{cccc}
    1 & 1 & \cdots & 1 \\
    x_1 & x_2 & \cdots & x_n \\
    \vdots & \vdots & \ddots & \vdots \\
    x_1^{d-2} & x_2^{d-2} & \cdots & x_n^{d-2} \\
  \end{array}
\right).\]

We claim that this coloring is proper. For any two distinct permutations $\sigma$ and $\pi$ with a same color $v\in \mathbb{Z}_p^{d-1}$, we have $A\sigma\equiv A\pi\equiv v \pmod{p}$. So $A(\sigma-\pi)\equiv0\pmod{p}$. Suppose the distance between $\sigma$ and $\pi$ is less than $d$, then there are at most $d-1$ nonzero coordinates in $\sigma-\pi$. Then we can deduce that there exist $d-1$ columns in $A$ which are linearly dependent in $\mathbb{Z}_p$, which is a contradiction to the fact that every $d-1$ columns in $A$ are linearly independent in $\mathbb{Z}_p$. Thus every two vertices with the same color are nonadjacent in $G_H$. So our coloring is proper.

Now we count the number $T$ of colors we used. The colors are in $\mathbb{Z}_p^{d-1}$ and note that the first coordinate is a constant $1+2+\dots+n\pmod{p}$. Thus $T\leq p^{d-2}$. Now each color corresponds to an independent set in $G_H$, or equivalently, an $(n,d)$-permutation code under Hamming metric. By Lemma \ref{color} we have
\[
|A_H(n,d)|\geq\frac{n!}{p^{d-2}}.
\]\qed

Consider the asymptotic behavior of our lower bound. The following notations simplify the upcoming statements and comparisons. In the remaining part of this section, $A_H(n,d)$ denotes the bound we get in Theorem \ref{pc}, $A_{H}^{GV}(n,d)$ denotes the classical Gilbert-Varshamov bound and $\widetilde{A}_H(n,d)$ denotes the lower bound derived in \cite{GYG}.

\begin{corollary}
When $d$ is a fixed constant and $n$ goes into infinity, $A_H(n,d)$ improves the classical Gilbert-Varshamov bound by a factor of $n$, that is,
\[
\frac{A_H(n,d)}{A_H^{GV}(n,d)}=\Omega(n).
\]
\end{corollary}

\pf
Since $D_k=\lfloor\frac{k!}{e}+\frac{1}{2}\rfloor$, we have
\[
B_H(d-1)=\sum_{k=0}^{d-1}|D(n,k)|=\sum_{k=0}^{d-1}{n \choose k}D_k=\Theta(n^{d-1}).
\]
It is well known \cite{prime} that there exists a prime $p$, satisfying $n\leq p\leq 2n$,
\[
A_H(n,d)\geq \frac{n!}{p^{d-2}}\geq \frac{n!}{(2n)^{d-2}}.
\]
Then
\[
\frac{A_H(n,d)}{A_H^{GV}(n,d)}\geq \frac{B_H(d-1)}{(2n)^{d-2}}=\Omega(n).
\]\qed

Furthermore, our lower bound also performs quite well when $n$ is small. For $d=5$ and $8\leq n\leq 20$, we list the results of $\widetilde{A}_H(n,d)$ and $A_H(n,d)$ in Table \ref{tbl}. Relatively better values are in bold form.

\begin{small}
\begin{table}[!t]
\centering
\caption{A comparison of $A_H(n,d)$ and $\widetilde{A}_H(n,d)$ with $d=5$ and $8 \le n \le 20$}
\label{tbl}

\begin{tabular}{c|cc}
  $n$ & $A_H(n,d)$ & $\widetilde{A}_H(n,d)$ \\
  \hline
  $8$ & $30$ & $\textbf{90}$ \\
  $9$ & $272$ & $\textbf{509}$ \\
  $10$ & $2726$ & $\textbf{3386}$ \\
  $11$ & $\textbf{29990}$ & $25885$ \\
  $12$ & $218025$ & $\textbf{223378}$ \\
  $13$ & $\textbf{2834328}$ & $2147724$ \\
  $14$ & $17744410$ & $\textbf{22767826}$ \\
\end{tabular}
~~~~
\begin{tabular}{c|cc}
  $n$ & $A_H(n,d)$ & $\widetilde{A}_H(n,d)$ \\
  \hline
  $15$ & $\textbf{266166164}$ & $263832788$ \\
  $16$ & $\textbf{4258658637}$ & $3317928906$ \\
  $17$ & $\textbf{72397196844}$ & $45006297715$ \\
  $18$ & $\textbf{933426695688}$ & $655021291542$ \\
  $19$ & $\textbf{17735107218083}$ & $10181693092799$ \\
  $20$ & $\textbf{199959070286565}$ & $168351610362186$ \\
  \\
\end{tabular}
\end{table}
\end{small}
To sum up, the analysis above gives:
\begin{theorem}
When $d$ is a fixed constant and $n$ goes into infinity,
\[
A_H(n,d)\geq \Omega(n\frac{n!}{B_H(d-1)}).
\]
\end{theorem}

A final remark is a comparison of our result with Proposition \ref{TVV}, the result obtained by Tait, Vardy and Verstra\"{e}te \cite{TVV}. They are restricted to the case when $d/n$ is a fixed ratio with $0<d/n<0.5$. Whereas our result considers the case when $d$ is fixed and $n$ goes into infinity, that is, the ratio $d/n$ goes into zero. So in some sense our result works
as a complement of theirs.

\section{A lower bound of permutation codes under Kendall's $\tau$-metric} \label{kendall}

In the rest of this paper we turn our attention into Kendall's $\tau$-metric. As has been noted in Proposition \ref{Barg}, the lower bound of $A_{K}(n,d)$ derived by \cite{Barg} meets the sphere-packing upper bound asymptotically for any fixed $d$. There's only a constant gap between the lower and upper bounds. In this section we attempt to narrow this gap.

For a permutation $\pi\in S_n$, an {\it inversion vector} $x_{\pi}=(x_{\pi}(2),x_{\pi}(3),\dots,x_{\pi}(n))\in \mathbb{Z}_n!\triangleq \mathbb{Z}_2 \times \mathbb{Z}_3 \times \dots \times \mathbb{Z}_n$ is defined as:
\begin{equation*}
x_{\pi}(i)=|\{j:j<i,\pi^{-1}(j)>\pi^{-1}(i)\}|, 2\le i \le n.
\end{equation*}
That is, $x_{\pi}(i)\in\mathbb{Z}_{i}$ counts the number of inversions formed by `$i$' and `$j$', $1\le j \le i-1$. For example, let $\pi=[4,5,2,1,3]$, then $x_{\pi}=(1,0,3,3)$. The sum of all entries equals the total number of inversions $I(\pi)$.

An adjacent transposition results in an error of weight one in the inversion vector $x_{\pi}$. Specifically, suppose we have two consecutive numbers $a$ and $b$ in the original permutation, with $a<b$. Then an adjacent transposition which switches $a$ and $b$ will result in an error $\mathbf{e^{+}_{b}}$, which is a vector with $+1$ on the entry $x_{\pi}(b)$ and $0$ elsewhere. Continuing the example, switch `$4$' and `$5$' in $\pi$, we have $\pi'=[5,4,2,1,3]$ and $x_{\pi'}=(1,0,3,4)$. Then $x_{\pi'}-x_{\pi}=(0,0,0,1)=\mathbf{e^{+}_{5}}$.

In contrast, if we have two consecutive numbers $b$ and $a$ in the original permutation, with $b>a$, then an adjacent transposition which switches $b$ and $a$ will result in an error $\mathbf{e^{-}_{b}}$, which is a vector with $-1$ on the entry $x_{\pi}(b)$ and $0$ elsewhere. Continuing the example, switch `$5$' and `$2$' in $\pi$, we have $\pi'=[4,2,5,1,3]$ and $x_{\pi'}=(1,0,3,2)$. Then $x_{\pi'}-x_{\pi}=(0,0,0,-1)=\mathbf{e^{-}_{5}}$.

Between two permutations, $t$ adjacent transpositions together lead to an error vector $\mathbf{e}$, which is the summation of each error vector corresponding to each adjacent transposition. Let $\omega(\mathbf{e})$ be the summation of all the entries in $\mathbf{e}$, performed over the integers. Note that since there may be some offsets of the form $\mathbf{e^{+}_{b}}$ and $\mathbf{e^{-}_{b}}$, $\omega(\mathbf{e})$ will be an integer with absolute value no more than $t$.

The key tool is the following famous theorem of Bose and Chowla \cite{Bose}:
\begin{lemma}{\rm\cite{Bose}} \label{Bose1}
Let $q$ be a power of a prime and $m=\frac{q^{t+1}-1}{q-1}$. There exist $q+1$ integers $d_1=0$, $d_2,\dots,d_{q+1}$ in $\mathbb{Z}_m$ such that the sums
\begin{equation*}
d_{i_1}+d_{i_2}+\dots+d_{i_t} ~~~~~~~~ (1\le i_1\le i_2\le\dots\le i_t\le q+1)
\end{equation*}
are all distinct modulo $m$.
\end{lemma}

Set $q+1=n-1$. We now deal with $A_{K}(n,2t+1)$. Color each permutation in $S_n$ using colors $(c_1,c_2)\in \mathbb{Z}_{2t+1}\times \mathbb{Z}_m$.

\begin{theorem} \label{colorkendall}
Under the parameters given above, for any permutation $\pi\in S_n$, let $c_1(\pi)\equiv I(\pi) \pmod{2t+1}$ and let $c_2(\pi)\equiv \sum_{i=1}^{n-1}d_i x_{\pi}(i+1) \pmod{m}$. Then for any two permutations $\pi$ and $\sigma$ with $d_{K}(\pi,\sigma)<2t+1$, we have $(c_1(\pi),c_2(\pi))\neq(c_1(\sigma),c_2(\sigma))$.
\end{theorem}

\pf
Let $\mathbf{e}=x_{\pi}-x_{\sigma}$ be the error vector between the inversion vectors of the two permutations. Since $d_{K}(\pi,\sigma)<2t+1$, $|\omega(\mathbf{e})|\le 2t$. If $|\omega(\mathbf{e})|\neq 0$, then clearly $c_1(\pi)\neq c_1(\sigma)$. Otherwise, $\omega(\mathbf{e})=0$, then the value of $c_2(\pi)-c_2(\sigma)$ modulo $m$ is the difference of two parts of summations. Each summation is the sum of some $s$ integers among $\{d_1,\dots,d_{q+1}\}$, with $s\le t$. By the Bose-Chowla theorem, this difference is nonzero so $c_2(\pi)\neq c_2(\sigma)$.\qed

So the coloring is a proper coloring in the graph $G_K$. A remark is that the proof of Barg and Mazumdar \cite{Barg} could be stated similarly in the framework above. Their coloring scheme aims at simultaneously dealing with all the possible error vectors. However, this simultaneousness also restricts the number of colors to be of order $t^2m$ (see Proposition \ref{Barg}), which is larger than $(2t+1)m$ in our approach. The trick in our framework, which deals with errors respectively according to whether $|\omega(\mathbf{e})|$ is zero or not, turns out to be useful. In summary, our coloring framework gives:

\begin{theorem}
Let $m=((n-2)^{t+1}-1)/(n-3)$, where $n-2$ is a prime power. Then $A_{K}(n,2t+1)\ge\frac{n!}{(2t+1)m}$.
\end{theorem}

As for $A_K(n,2t)$, by the theorem above and Lemma \ref{even}, we immediately have:

\begin{theorem}
Let $m=((n-2)^{t}-1)/(n-3)$, where $n-2$ is a prime power. Then $A_{K}(n,2t)\ge\frac{n!}{2(2t-1)m}$.
\end{theorem}

We mention that there's still a slight chance of doing better. In our framework above, when dealing with errors with $|\omega(\mathbf{e})|\neq 0$, we calculate the inversion number of a permutation modulo $2t+1$. Could we lower this number? We now state another Bose-Chowla theorem also appeared in \cite{Bose}. Note that in the framework above, both of the two Bose-Chowla theorems could be applied, and actually they lead to similar results, with Lemma \ref{Bose1} performing slightly better. However, the following Lemma \ref{Bose2} benefits our analysis later.

\begin{lemma}{\rm\cite{Bose}} \label{Bose2}
Let $q=p^n$ be a prime power. Then we can find $q$ nonzero integers (less than $q^t$) $d_1=1,d_2,\dots,d_q$ such that the sums
\begin{equation*}
d_{i_1}+d_{i_2}+\dots+d_{i_t} ~~~~~~~~ (1\le i_1\le i_2\le\dots\le i_t\le q)
\end{equation*}
are all distinct modulo $q^t-1$.
\end{lemma}

The exact constructions of these integers are as follows. Let $\alpha_1=0,\alpha_2,\dots,\alpha_q$ denote all the elements in $\mathbb{F}_{p^n}$. Let $y$ be a primitive element of the extension field $\mathbb{F}_{p^{nt}}$. Let $y^{d_i}=y+\alpha_i$ for $i=1,2\dots,q$, where $d_i<p^{nt}$. Then $\{d_i\}_{1\le i \le q}$ is the desired set of integers for carrying out our coloring scheme in Theorem \ref{colorkendall}. The choice of the primitive element $y$, or equivalently, the choice of the irreducible polynomial of degree $t$ with coefficients from $\mathbb{F}_{p^n}$, uniquely determines $\{d_i\}_{1\le i \le q}$. We now expect more properties from the choice of the irreducible polynomial.

Take $A_{K}(n,5)$ as an example. Now we need an appropriate irreducible polynomial of degree $2$ with coefficients from $\mathbb{F}_{p^n}$, denoted as $y^2=ay+b$, $a,b\in \mathbb{F}_{p^n}$. We further demand that the set of integers $\{d_i\}_{1\le i \le q}$ satisfies: the sum of any three integers is nonzero modulo $p^{2n}-1$. That is, $(y+i)(y+j)(y+k)\neq1$ for any $i,j,k\in \mathbb{F}_{p^n}$. It can be checked that this is equivalent to the following problem.

\textbf{Problem}: For any prime power $p^n$, find $a,b\in \mathbb{F}_{p^n}$ such that

$\bullet$ $y^2=ay+b$ is an irreducible polynomial in $\mathbb{F}_{p^n}$, and

$\bullet$ the following system of equations,
\begin{equation*}
\left\{
\begin{aligned}
&a^2+b+ai+aj+ak+ij+ik+jk=0 \\
& ab+ib+jb+kb+ijk=1 \\
\end{aligned}
\right.
\end{equation*}
with $i,j,k$ being indeterminate, has no solution in $\mathbb{F}_{p^n}^3$.

Via computer search, although the desired $a$ and $b$ do not exist for $\mathbb{F}_{5}$ and $\mathbb{F}_{7}$, yet they do exist for primes 11, 13, 17, 19, 23. We conjecture that it may be true that there are infinitely many prime powers for which the desired $a$ and $b$ exist.

Once $a$ and $b$ exist for a prime power $p^n$, then we could do a small adjustment for our coloring map. Now let $\widetilde{c}_1(\pi)\equiv I(\pi) \pmod{3}$ instead of modulo $5$. For any two permutations $\sigma$ and $\tau$ with $d_{K}(\sigma,\pi)<5$, the only possibility for $\widetilde{c}_1(\sigma)=\widetilde{c}_1(\pi)$ and $\omega(x_{\pi}-x_{\sigma})\neq0$ is that the error vector between their inversion vectors $x_{\sigma}$ and $x_{\pi}$ is a vector with exactly three entries being `1' and otherwise `0'. Then the difference of $c_2(\sigma)$ and $c_2(\pi)$ will be a summation of three integers out of $\{d_i\}_{1\le i \le q}$. But by the further demand of the properties of $a$ and $b$ we choose, this difference is ensured to be nonzero modulo $p^{2n}-1$. Thus it is guaranteed that $c_2(\sigma)\neq c_2(\pi)$. So in this manner, the constant gap between the lower bound and the sphere-packing upper bound could be a little bit smaller.

\section{Sporadic results on $A_{K}(n,d)$} \label{sporadic}

In this section we provide some other sporadic results concerning permutation codes under Kendall's $\tau$-metric.

\subsection{A generalization of the code-anticode method}

First we take a look at Lemma \ref{local}. The code-anticode method used in \cite{Buzaglo2} corresponds to finding a subset $G'$ of vertices, satisfying $\alpha(G')=1$, with $|G'|$ as large as possible. A natural generalization is to jump out of the restriction $\alpha(G')=1$. That is, as Lemma \ref{local} suggests, we want to search for a subset $G'$ with $\alpha(G')/|G'|$ as small as possible. An illustrative example is the following precise determination of the value $A_{K}(5,3)$. In \cite{Buzaglo2} it has been verified that $20\le A_{K}(5,3) \le 23$. We now show that 20 is the exact value.

\begin{theorem}
$$A_K(5,3)=20.$$
\end{theorem}

\pf
Select $G'=\{[1,2,3,4,5]$, $[1,2,3,5,4]$, $[1,2,4,3,5]$, $[1,2,4,5,3]$, $[1,2,5,3,4]$, $[1,2,5,4,3]$, $[2,1,3,4,5]$, $[2,1,3,5,4]$, $[2,1,4,3,5]$, $[2,1,4,5,3]$, $[2,1,5,3,4]$, $[2,1,5,4,3]\}$. It can be easily verified that $\alpha(G')=2$. So we have $\frac{A_{K}(5,3)}{5!}\le \frac{\alpha(G')}{|G'|}=\frac{2}{12}$, which leads to $A_{K}(5,3)\le 20$ and thus fixes this value.\qed

Although this is only a simple case, yet the idea lying behind it may have potentials for other parameters, or even perhaps for analyzing upper bounds for other various codes.

\subsection{Sporadic results by computer search}

Some other sporadic results concerning small parameters $n=5$ and $n=6$ could be obtained by computer search, via some algorithms designed for searching maximal independent sets. We obtain some values better than those listed in the table in \cite{Buzaglo2}, by the program developed by Ashay Dharwadker \cite{algorithm}. These values are listed as follows and their corresponding codewords are listed in the appendix. Although lacking strictly mathematical analysis, the power of the program suggests that these may be the exact values.
\begin{theorem}
\begin{align*}
&A_{K}(5,4)\ge 12, A_{K}(5,6)\ge 5, \\
&A_{K}(6,3)\ge 101, A_{K}(6,4)\ge 64, A_{K}(6,5)\ge 25, \\
&A_{K}(6,6)\ge 20, A_{K}(6,7)\ge 11, A_{K}(6,8)\ge 7. \\
\end{align*}
\end{theorem}

\subsection{Counting pairs of inversions: a Plotkin-type bound}

In this subsection we prove a Plotkin-type bound by counting pairs of inversions. Recall that we have the following expression for Kendall's $\tau$-metric: $$d_K(\sigma,\pi)=|\{ (i,j): \sigma^{-1}(i)<\sigma^{-1}(j) \wedge \pi^{-1}(i)>\pi^{-1}(j)\}|.$$

\begin{theorem} \label{count}
If $A_K(n,2t)\ge M$, then
$$2{M\choose 2}t \le {n\choose 2}\lceil{\frac{M}{2}}\rceil\lfloor{\frac{M}{2}}\rfloor,$$
and if $A_K(n,2t+1)\ge M$, then
$${{\lceil{\frac{M}{2}}\rceil}\choose 2}(2t+2)+{{\lfloor{\frac{M}{2}}\rfloor}\choose 2}(2t+2)+\lceil{\frac{M}{2}}\rceil\lfloor{\frac{M}{2}}\rfloor(2t+1) \le {n\choose 2}\lceil{\frac{M}{2}}\rceil\lfloor{\frac{M}{2}}\rfloor.$$
\end{theorem}

\pf
Suppose now we have an $(n,d)$-permutation code $\mathcal{C}$ under Kendall's $\tau$-metric of size $M$. We now calculate the summation of all the pair-wise distances $\sum=\Sigma_{c_1,c_2\in \mathcal{C}} d_K(c_1,c_2)$. Firstly, for any pair of numbers $1\le i < j\le n$, we could partition $\mathcal{C}$ into two parts, according to whether $i$ precedes $j$ or vice versa. Then from the expression for Kendall's $\tau$-metric, we know that the pair $(i,j)$ contributes one to the distance between two permutations from different parts. Thus, the pair $(i,j)$ contributes at most $\lceil\frac{M}{2}\rceil\lfloor\frac{M}{2}\rfloor$ to $\sum$. So we have,

$$\sum\le {n\choose 2}\lceil{\frac{M}{2}}\rceil\lfloor{\frac{M}{2}}\rfloor.$$

On the other hand, $\sum\ge {M\choose 2}d$. And especially, if the distance $d$ is odd, since the Kendall's $\tau$-distance between two permutations of the same parity is even, then $\sum\ge {{\lceil{\frac{M}{2}}\rceil}\choose 2}(d+1)+{{\lfloor{\frac{M}{2}}\rfloor}\choose 2}(d+1)+\lceil{\frac{M}{2}}\rceil\lfloor{\frac{M}{2}}\rfloor d$.

The theorem follows from a comparison of the upper bound and lower bound of $\sum$.\qed

Now we analyze when will the theorem above be useful. Using the first constraint as an example, when $d\le \frac{1}{2}{n\choose 2}$, the constraint naturally holds for any $M$ and thus does not provide any useful bound on $M$. However, when $\frac{1}{2}{n\choose 2}< d\le \frac{2}{3}{n\choose 2}$, this bound may turn out to be better than the sphere packing upper bound or Proposition \ref{BEEven}. It is generally unrealistic to compare the Plotkin-type bound to the sphere-packing upper bound or Proposition \ref{BEEven} since the precise size of a Kendall's $\tau$-ball or a double-ball is difficult to analyze. Below we list several cases for small parameters, as supporting evidences to show that Theorem \ref{count} may work slightly better.

\begin{small}
\begin{table}[!htb]
\begin{center}
\begin{tabular}{|c|c|c|c|}
  \hline
  $n$ & $d$ & Sphere-packing bound & Theorem \ref{count} \\\hline
  6 & 9 & 7 & 4 \\\hline
  7 & 13 & 8 & 4 \\\hline
  7 & 11 & 14 & 12 \\\hline
  8 & 17 & 11 & 4 \\\hline
\end{tabular}
~~~~
\begin{tabular}{|c|c|c|c|}
  \hline
  $n$ & $d$ & Proposition \ref{BEEven} & Theorem \ref{count} \\\hline
  7 & 12 & 11 & 8 \\\hline
  8 & 18 & 9 & 4 \\\hline
  8 & 16 & 14 & 8 \\\hline
   &  &  &  \\\hline
\end{tabular}
\end{center}
\end{table}
\end{small}

\section{Conclusions} \label{conclusion}
Permutation codes under different metrics are interesting topics due to their various applications. The bounds of permutation codes can be analyzed via studying the independence numbers of the corresponding graphs. We use a coloring approach to analyze the independence numbers, which leads to some improvements on the lower bounds of permutation codes under Hamming metric and Kendall's $\tau$-metric, respectively. Although this coloring approach is well-known, the tricky part is the coloring strategy case-by-case. Besides, we also derive some other sporadic results concerning the upper bound of permutation codes under Kendall's $\tau$-metric.

\bibliographystyle{plain}
\bibliography{ref2}

\begin{thebibliography}{10}

\bibitem{Barg}
A.~Barg and A.~Mazumdar.
\newblock Codes in permutations and error correction for rank modulation.
\newblock {\em IEEE Trans. Inform. Theory}, 56(7):3158--3165, 2010.

\bibitem{Bose}
R.~C. Bose and S.~Chowla.
\newblock Theorems in the additive theory of numbers.
\newblock {\em Comment. Math. Helv.}, 37:141--147, 1962/1963.

\bibitem{Buzaglo2}
S.~Buzaglo and T.~Etzion.
\newblock Bounds on the size of permutation codes with the {K}endall
  {$\tau$}-metric.
\newblock {\em IEEE Trans. Inform. Theory}, 61(6):3241--3250, 2015.

\bibitem{prime}
P.~L. {\v{C}}eby{\v{s}}ev.
\newblock {\em M{\'e}moire sur les nombres premiers}.
\newblock Acad{\'e}mie Imp{\'e}riale des Sciences, 1850.

\bibitem{CCD}
W.~Chu, C.~J. Colbourn, and P.~Dukes.
\newblock Constructions for permutation codes in powerline communications.
\newblock {\em Des. Codes Cryptogr.}, 32(1-3):51--64, 2004.

\bibitem{CKL}
C.~J. Colbourn, T.~Kl{\o}ve, and A.~C.~H. Ling.
\newblock Permutation arrays for powerline communication and mutually
  orthogonal {L}atin squares.
\newblock {\em IEEE Trans. Inform. Theory}, 50(6):1289--1291, 2004.

\bibitem{DCL}
D.~R. de~la Torre, C.~J. Colbourn, and A.~C.~H. Ling.
\newblock An application of permutation arrays to block ciphers.
\newblock {\em Congr Numer}, pages 5--7, 2000.

\bibitem{algorithm}
A.~Dharwadker.
\newblock {\em The independent set algorithm}.
\newblock 2006.

\bibitem{Farnoud}
F.~Farnoud, V.~Skachek, and O.~Milenkovic.
\newblock Error-correction in flash memories via codes in the {U}lam metric.
\newblock {\em IEEE Trans. Inform. Theory}, 59(5):3003--3020, 2013.

\bibitem{FV}
H.~C. Ferreira and A.~J.~H. Vinck.
\newblock Interference cancellation with permutation trellis codes.
\newblock In {\em Vehicular Technology Conference, 2000. IEEE-VTS Fall VTC
  2000. 52nd}, pages 2401--2407 vol.5, 2000.

\bibitem{FD}
P.~Frankl and M.~Deza.
\newblock On the maximum number of permutations with given maximal or minimal
  distance.
\newblock {\em J. Combinatorial Theory Ser. A}, 22(3):352--360, 1977.

\bibitem{GYG}
F.~Gao, Y.~Yang, and G.~Ge.
\newblock An improvement on the {G}ilbert-{V}arshamov bound for permutation
  codes.
\newblock {\em IEEE Trans. Inform. Theory}, 59(5):3059--3063, 2013.

\bibitem{Godsil}
C.~Godsil and G.~Royle.
\newblock {\em Algebraic graph theory}, volume 207 of {\em Graduate Texts in
  Mathematics}.
\newblock Springer-Verlag, New York, 2001.

\bibitem{Jiang}
A.~Jiang, R.~Mateescu, M.~Schwartz, and J.~Bruck.
\newblock Rank modulation for flash memories.
\newblock {\em IEEE Trans. Inform. Theory}, 55(6):2659--2673, 2009.

\bibitem{Jiang2}
A.~Jiang, M.~Schwartz, and J.~Bruck.
\newblock Correcting charge-constrained errors in the rank-modulation scheme.
\newblock {\em IEEE Trans. Inform. Theory}, 56(5):2112--2120, 2010.

\bibitem{Klove}
T.~Kl{\o}ve, T.-T. Lin, S.-C. Tsai, and W.-G. Tzeng.
\newblock Permutation arrays under the {C}hebyshev distance.
\newblock {\em IEEE Trans. Inform. Theory}, 56(6):2611--2617, 2010.

\bibitem{Mazumdar}
A.~Mazumdar, A.~Barg, and G.~Z{\'e}mor.
\newblock Constructions of rank modulation codes.
\newblock {\em IEEE Trans. Inform. Theory}, 59(2):1018--1029, 2013.

\bibitem{PVYH}
N.~Pavlidou, A.~J.~H. Vinck, J.~Yazdani, and B.~Honary.
\newblock Power line communications: State of the art and future trends.
\newblock {\em IEEE Communications Magazine}, 41(4):34--40, 2003.

\bibitem{TVV}
M.~Tait, A.~Vardy, and J.~Verstraete.
\newblock Asymptotic improvement of the gilbert-varshamov bound on the size of
  permutation codes.
\newblock {\em arXiv preprint arXiv:1311.4925}, 2013.

\bibitem{Tamo}
I.~Tamo and M.~Schwartz.
\newblock Correcting limited-magnitude errors in the rank-modulation scheme.
\newblock {\em IEEE Trans. Inform. Theory}, 56(6):2551--2560, 2010.

\bibitem{V}
A.~J.~H. Vinck.
\newblock Coded modulation for power line communications.
\newblock In {\em AE Int. J. Electron. and Commun}, pages 45--49, 2011.

\end{thebibliography}

\section*{Appendix}

~~~~~A (5,4)-permutation code under Kendall's $\tau$-metric with 12 codewords:
\begin{align*}
& [1,2,3,4,5], [1,3,5,4,2], [1,4,5,2,3], [2,1,5,4,3], [2,4,3,1,5], [3,4,5,1,2], \\
& [3,5,2,1,4], [4,1,3,2,5], [4,2,5,1,3], [5,1,2,3,4], [5,2,4,3,1], [5,4,1,3,2]. \\
\end{align*}

A (5,6)-permutation code under Kendall's $\tau$-metric with 5 codewords:
\begin{equation*}
  [1,2,3,4,5], [2,4,5,3,1], [3,5,2,1,4], [4,3,1,5,2], [5,1,4,2,3]. \\
\end{equation*}

A (6,3)-permutation code under Kendall's $\tau$-metric with 101 codewords:
\begin{align*}
& [1,2,3,4,6,5],  [1,2,5,4,3,6],  [1,2,6,4,3,5],  [1,2,6,5,3,4],  [1,3,2,6,5,4],  [1,3,4,6,2,5],\\
& [1,3,5,2,4,6],  [1,3,6,5,4,2],  [1,4,2,5,6,3],  [1,4,3,2,5,6],  [1,4,5,6,3,2],  [1,4,6,2,3,5],\\
& [1,5,2,3,6,4],  [1,5,3,4,6,2],  [1,5,6,4,2,3],  [1,6,4,3,5,2],  [1,6,5,3,2,4],  [2,1,3,5,6,4],\\
& [2,1,4,3,5,6],  [2,1,4,6,5,3],  [2,3,1,6,4,5],  [2,3,4,1,5,6],  [2,3,6,4,5,1],  [2,4,3,5,6,1],\\
& [2,4,5,1,6,3],  [2,4,6,1,3,5],  [2,5,1,6,4,3],  [2,5,3,1,6,4],  [2,5,4,3,1,6],  [2,5,6,3,4,1],\\
& [2,6,1,3,4,5],  [2,6,1,5,4,3],  [2,6,3,5,1,4],  [2,6,4,5,1,3],  [3,1,2,4,5,6],  [3,1,4,5,6,2],\\
& [3,1,5,6,2,4],  [3,1,6,2,4,5],  [3,2,4,6,1,5],  [3,2,5,1,4,6],  [3,2,5,6,4,1],  [3,2,6,1,5,4],\\
& [3,4,1,2,6,5],  [3,4,2,5,6,1],  [3,4,5,1,2,6],  [3,4,6,5,1,2],  [3,5,6,4,2,1],  [3,6,1,4,5,2],\\
& [3,6,4,2,1,5],  [3,6,5,2,1,4],  [4,1,2,3,6,5],  [4,1,3,6,5,2],  [4,1,5,3,2,6],  [4,1,6,5,2,3],\\
& [4,2,1,5,3,6],  [4,2,3,6,1,5],  [4,2,6,5,3,1],  [4,3,2,1,5,6],  [4,3,5,6,2,1],  [4,3,6,1,2,5],\\
& [4,5,1,2,6,3],  [4,5,2,3,6,1],  [4,5,3,1,6,2],  [4,5,6,2,1,3],  [4,6,2,1,5,3],  [4,6,3,2,5,1],\\
& [4,6,5,1,3,2],  [5,1,2,4,6,3],  [5,1,4,3,2,6],  [5,1,6,2,3,4],  [5,2,1,3,4,6],  [5,2,3,4,6,1],\\
& [5,2,4,6,1,3],  [5,2,6,1,3,4],  [5,3,1,2,4,6],  [5,3,1,6,4,2],  [5,3,2,6,1,4],  [5,3,4,2,1,6],\\
& [5,3,4,6,1,2],  [5,4,1,6,3,2],  [5,4,2,1,3,6],  [5,4,6,3,2,1],  [5,6,1,3,4,2],  [5,6,3,2,4,1],\\
& [5,6,4,1,2,3],  [6,1,2,3,5,4],  [6,1,2,4,5,3],  [6,1,3,4,2,5],  [6,1,5,4,3,2],  [6,2,4,3,1,5],\\
& [6,2,5,1,3,4],  [6,2,5,4,3,1],  [6,3,1,5,2,4],  [6,3,2,1,4,5],  [6,3,2,5,4,1],  [6,3,4,5,2,1],\\
& [6,4,1,2,3,5],  [6,4,3,1,5,2],  [6,4,5,2,3,1],  [6,5,1,2,4,3],  [6,5,4,3,1,2].
\end{align*}

A (6,4)-permutation code under Kendall's $\tau$-metric with 64 codewords:
\begin{align*}
& [1,2,3,5,6,4],  [1,2,4,6,5,3],  [1,3,2,4,6,5],  [1,3,4,5,6,2],  [1,4,2,3,5,6],  [1,5,2,6,4,3],\\
& [1,5,4,3,2,6],  [1,5,6,3,4,2],  [1,6,2,5,3,4],  [1,6,3,4,2,5],  [1,6,4,5,2,3],  [2,1,5,4,3,6],\\
& [2,1,6,3,4,5],  [2,3,1,4,5,6],  [2,3,5,6,1,4],  [2,3,6,4,1,5],  [2,4,1,3,6,5],  [2,4,6,5,1,3],\\
& [2,5,1,6,3,4],  [2,5,4,3,6,1],  [2,6,1,5,4,3],  [2,6,5,3,4,1],  [3,1,5,2,4,6],  [3,1,6,5,4,2],\\
& [3,2,1,6,5,4],  [3,2,5,4,1,6],  [3,4,2,1,6,5],  [3,4,5,1,2,6],  [3,5,4,6,2,1],  [3,5,6,1,2,4],\\
& [3,6,1,2,4,5],  [3,6,2,5,4,1],  [3,6,4,5,1,2],  [4,1,5,6,3,2],  [4,1,6,2,3,5],  [4,2,1,5,6,3],\\
& [4,2,3,5,1,6],  [4,2,6,3,1,5],  [4,3,1,6,5,2],  [4,3,6,2,5,1],  [4,5,1,2,3,6],  [4,5,2,6,3,1],\\
& [4,5,3,6,1,2],  [4,6,5,1,2,3],  [5,1,3,2,6,4],  [5,1,4,6,2,3],  [5,2,3,1,4,6],  [5,2,4,1,6,3],\\
& [5,3,1,4,6,2],  [5,3,2,6,4,1],  [5,4,3,2,1,6],  [5,6,1,2,3,4],  [5,6,3,4,1,2],  [5,6,4,2,1,3],\\
& [6,1,2,4,3,5],  [6,1,3,5,2,4],  [6,2,3,1,5,4],  [6,2,4,3,5,1],  [6,3,4,2,1,5],  [6,4,1,3,5,2],\\
& [6,4,2,1,5,3],  [6,4,5,3,2,1],  [6,5,1,4,3,2],  [6,5,3,2,1,4].
\end{align*}

A (6,5)-permutation code under Kendall's $\tau$-metric with 25 codewords:
\begin{align*}
& [1,2,3,4,6,5],  [1,3,5,4,2,6],  [1,5,4,6,2,3],  [1,6,3,5,2,4],  [2,1,4,5,6,3],  [2,3,6,4,5,1],\\
& [2,5,4,3,6,1],  [2,6,1,5,3,4],  [3,1,6,4,2,5],  [3,2,1,5,6,4],  [3,5,4,2,6,1],  [4,1,3,6,5,2],\\
& [4,2,3,1,5,6],  [4,2,6,5,1,3],  [4,3,6,2,5,1],  [4,5,1,2,3,6],  [5,2,1,3,4,6],  [5,3,1,6,4,2],\\
& [5,4,6,3,1,2],  [5,6,1,2,3,4],  [6,1,4,5,3,2],  [6,2,4,1,3,5],  [6,3,2,5,1,4],  [6,3,4,5,1,2],\\
& [6,5,2,4,3,1].
\end{align*}

A (6,6)-permutation code under Kendall's $\tau$-metric with 20 codewords:
\begin{align*}
& [1,2,3,4,6,5],  [1,5,4,3,6,2],  [1,6,3,5,2,4],  [1,6,4,2,5,3],  [2,1,5,4,6,3],  [2,3,4,5,6,1],\\
& [2,6,4,1,3,5],  [2,6,5,3,1,4],  [3,2,1,5,6,4],  [3,4,5,1,6,2],  [3,6,1,4,2,5],  [3,6,5,2,4,1],\\
& [4,3,2,1,6,5],  [4,5,1,2,6,3],  [4,6,1,3,5,2],  [4,6,2,5,3,1],  [5,1,2,3,6,4],  [5,4,3,2,6,1],\\
& [5,6,2,4,1,3],  [5,6,3,1,4,2].
\end{align*}

A (6,7)-permutation code under Kendall's $\tau$-metric with 11 codewords:
\begin{align*}
& [1,2,3,4,5,6],  [1,5,4,3,6,2],  [2,1,6,5,4,3],  [2,6,3,4,5,1],  [3,4,5,6,1,2],  [3,5,2,1,6,4],\\
& [4,3,2,1,6,5],  [4,5,2,1,6,3],  [5,6,1,2,3,4],  [6,1,3,4,2,5],  [6,5,4,3,2,1].
\end{align*}

A (6,8)-permutation code under Kendall's $\tau$-metric with 7 codewords:
\begin{equation*}
  [1,2,3,6,4,5],  [1,4,5,6,2,3],  [2,4,5,3,1,6],  [3,4,6,2,1,5],  [3,5,1,4,2,6],  [5,2,6,1,3,4],  [6,5,4,3,1,2].
\end{equation*}

\end{document}